\documentclass[11pt,oneside]{article}

\usepackage{graphicx}
\usepackage{amsmath}
\usepackage{amsfonts}
\usepackage{amssymb}
\usepackage{amsthm}
\usepackage{enumerate}
\usepackage{datetime}
\usepackage{url}
\usepackage{times} 
\usepackage[round,colon,authoryear]{natbib} 

\newcommand{\E}{\mathbb{E}}

\newcommand{\1}{\mathbf{1}}
\newcommand{\Prob}{\mathbb{P}}

\newcommand{\CF}{\mathcal{F}}

\newcommand{\dd}{\mathrm{d}}

\theoremstyle{remark}

\theoremstyle{remark}

\theoremstyle{remark}

\theoremstyle{remark}

\theoremstyle{remark}

\allowdisplaybreaks

\title{Negative Call Prices\thanks{I am grateful to Travis Fisher, Mike Hogan, Ioannis Karatzas, Arseniy Kukanov, Radka Pickova, Philip
Protter, Sergio Pulido,  Murad Taqqu, and Mike Tehranchi for fruitful discussions on the subject
matter of this paper. I thank an anonymous referee for her or his helpful comments. This work was partially supported by the
National Science Foundation DMS Grant
 09-05754.}}

\author{Johannes Ruf\footnote{E-mail: johannes.ruf@oxford-man.ox.ac.uk} \\Oxford-Man Institute of Quantitative Finance and Department of Mathematics\\ University of Oxford \\ 
\\
    \today
    }
\date{}

\begin{document}
\thispagestyle{plain} \maketitle
\begin{abstract}
\noindent
    We show that the existence of an equivalent local martingale measure for asset prices does not prevent negative prices for
    European calls written on positive stock prices.
    In particular, we illustrate that many standard no-arbitrage arguments implicitly rely on
    conditions stronger than the No Free Lunch With Vanishing Risk (NFLVR) assumption.  The discrepancy between replicating prices and market prices for a contingent claim may be observed in a model satisfying NFLVR since  
    certain trading strategies of buying one portfolio and selling another one are  often excluded by standard admissibility constraints.\\
\bf{Keywords}: Pricing; Hedging;  Arbitrage;  Admissibility;  NFLVR;  Contingent Claim;  Strict local martingale\\
\bf{JEL classification code}: G13
\end{abstract}
\section{Introduction}
    In the following, we illustrate how contingent claim prices can become negative, although the terminal payoff associated
    to the contingent claim is nonnegative.  The economy that we consider as an example satisfies the assumption of
    \emph{No Free Lunch With Vanishing Risk (NFLVR)}: In particular,
    there exists no (admissible) trading strategy that starts with zero initial wealth, has a wealth process uniformly bounded from below,  and leads to a terminal wealth
    that is
    always nonnegative and strictly positive with positive probability.  The NFLVR assumption in conjunction with local boundedness of asset prices 
    is equivalent to the existence of an \emph{equivalent local martingale measure (ELMM)}, to wit, a
    probability measure that is equivalent to the original one, and under which all asset price processes in the economy
    are
    local martingales.  For a precise statement of NFLVR and the proof of this equivalence, we refer the reader to
    \citet{DS_fundamental, DS_1998, DS_mono}.

    The following discussion involves market prices; these usually do not have to agree with replicating prices.
    This is illustrated by the existence of \emph{bubbles}, which are asset price processes whose current market price is higher
    than the costs for replicating them at some given time in the future; see \citet{JP_complete}, \citet{Ruf_hedging},
    and the discussion below.     In the economic literature, a bubble is sometimes interpreted as an asset that is overpriced,
    but nevertheless bought by agents since they believe
    that the asset can be sold in the future at an even higher price before the ``bubble bursts.''  Our example
    below can be interpreted similarly: It discusses an asset that is underpriced but nevertheless sold at the current price,
    which is lower than its intrinsic value, since the price might decrease even further in the future.
    To the best of our knowledge, models for an economic \emph{depression} have not been discussed in the framework
    of arbitrage-free pricing.

    It is not the purpose of this paper to make a case for the existence of negative call prices. On the contrary,
    we are convinced that negative call prices or, more generally, negative prices for contingent claims with positive
    terminal payoffs, should be excluded in an economy where agents prefer more to less.\footnote{Negative asset prices
    can, however, be observed in the market, for instance in the wind energy market. These negative
    prices occur primarily due to storage costs; see for the example the Bloomberg article
    \emph{Windmill Boom Cuts Electricity Prices in Europe} by J.~van Loon
    from April 23, 2010, retrieved from
    \url{http://www.bloomberg.com/news/2010-04-22/windmill-boom-curbs-electric-power-prices.html}. In this paper, we
    assume a frictionless market, in particular, an agent does not incur costs from holding an asset.}
    However, it is our aim to convey
    that many no-arbitrage arguments, such as the one showing the equality of American and European call prices for
    stocks that
    do not pay dividends, implicitly rely on stronger assumptions than just the existence of an ELMM.

	\subsubsection*{Admissibility constraints}
Describing the class of \emph{admissible} trading strategies, defined as trading strategies that an agent is allowed to follow, is essential for any formulation of a \emph{Fundamental Theorem of Asset Pricing} (FTAP) when asset price processes are exogenously given.  
An FTAP is usually formulated as the equivalence of the lack of an arbitrage opportunity and the existence of a certain probability measure, the so-called risk-neutral measure, under which asset prices have certain dynamics.  Towards this end, a precise definition of an arbitrage opportunity needs to be given.
Indeed, in any non-trivial
    infinite-horizon discrete-time or continuous-time model, such as the Black-Scholes model,
    notorious \emph{doubling strategies} exist, which lead to an arbitrage opportunity if they are not
    prohibited; see Section~6 of \citet{HarrisonKreps}.   
    
In order to avoid the trivial statement that an arbitrage opportunity exists in any continuous-time model, certain restrictions on the class of admissible trading strategies from which arbitrage opportunities may be selected have to be enforced. It is clear that the larger the class of admissible trading strategies is chosen, the stronger are the assumptions on the risk-neutral measure in order to have equivalence in the FTAP. In all cases we are aware of, admissible trading strategies are defined as the ones which lead to  wealth processes that are somehow bounded from below.
	
	 The classical approach, as suggested by \citet{HarrisonKreps} and  \citet{DS_fundamental}, is to require the wealth process to be uniformly bounded from below by a (negative) constant.  This can be motivated from
    an economic perspective as a margin requirement: As soon as an agent's (``she'') wealth
    reaches some specified negative
    wealth, her broker forces her to cancel her position and prevents her from
    further trading.  Under this admissibility constraint, no arbitrage (in the sense of NFLVR) corresponds to the existence of an equivalent probability measure, under which all asset price processes  follow local martingale dynamics given they are locally bounded.
    
    \citet{Yan_new} suggests to use a larger class of admissible trading strategies, namely the ones whose associated wealth process is bounded from below by a (negative) constant times the market portfolio. In particular, as the asset prices increase, the wealth process of an admissible trading strategy is allowed to become more and more negative.  As observed before, the extension of the class of admissible trading strategies implies a stronger no-arbitrage condition and thus leads to a risk-neutral measure satisfying a stronger condition; here, one under which all asset price processes follow true martingale dynamics. 
    
    The advantage of Yan's admissibility constraint is that it is independent of the choice of num\'eraire. Moreover, it excludes many pathologies such as the one studied here. However, its strong no-arbitrage assumptions exclude the possibility to model several important phenomena, such as bubbles as strict local martingales, relative arbitrage opportunities as in \citet{FK_survey}, or quadratic normal volatility models, which provide certain symmetry properties under a change of num\'eraire, as studied in \citet{CFR2011, CFR_qnv}.
	
	 Given an economy, under which asset price processes follow local martingale dynamics, for instance, it is interesting to extend the class of admissible  trading strategies without introducing arbitrage.  This was for example done in Proposition~4.1 of \citet{HLW} and more generally, in \citet{Strasser2003}, where a criterion is given on trading strategies, such that the corresponding wealth processes are supermartingales.

 After this discussion, the subtle reason for the existence of arbitrage-free models with  counter-intuitive price processes is clear.  A price might seem to imply an arbitrage opportunity but agents in the economy are not  permitted to profit from this ostensible arbitrage, due to admissibility constraints in their set of trading strategies.  More precisely, standard no-arbitrage arguments often
    imply the construction of a trading strategy consisting of selling one asset (for example, an European call)
    and buying another asset (for example, an American call).  It is implicitly utilized that such a trading strategy
    is admissible. Thus, this argument resembles more an assumption on the admissibility of a trading strategy than
    a clean no-arbitrage argument. 

    A related way to think about the existence of prices that seem to contradict simple no-arbitrage
    arguments is to study strict local
    martingales; that is, local martingales that are not martingales.  Any local martingale that is bounded from below
    by a constant
    is a supermartingale by Fatou's lemma; thus, any local martingale that is bounded from above by a constant
    is a submartingale.
    Therefore, if an asset price is modeled as a strict local martingale that is bounded from above, then the trading
    strategy of holding that asset for a fixed time is inadmissible, since its corresponding wealth process is not a supermartingale,
    but a submartingale. In the example below, we will make use of this insight.
    
    Indeed, the existence of \emph{bubbles}, modeled as positive strict local martingales, in models satisfying NFLVR
    is justified in the literature by the observation that selling such assets might represent an inadmissible
    trading strategy; see \citet{CH}, \citet{HLW}, and \citet{JP_complete}. Their argument can be marginally
    generalized by not restraining oneself to trading strategies that lead to a wealth process bounded from below, but
    by using the larger class of trading strategies discussed in \citet{Strasser2003}.

	\section{Example}
    In the following, we provide an example for an economy that satisfies NFLVR but allows for a negative call price.
    To this end, we fix a filtered probability space $(\Omega, \mathcal{F}, \{\CF(t)\}_{t \geq 0}, \Prob)$,
    where the filtration $\{\CF(t)\}_{t \geq 0}$ is generated by a Brownian motion $B(\cdot)$.
    We model an asset with initial price $S_1(0) = 1$ and price dynamics
    given as a geometric Brownian motion; that is,
    \begin{align*}
        \dd S_1(t) = S_1(t) \dd B(t)
    \end{align*}
    for all $t \geq 0$.

    We now consider an European at-the-money call with maturity $1$ written on $S_1(\cdot)$; to wit, we study
    an asset that at time $1$ pays precisely $D=(S_1(1) - 1)^+$, where $x^+$ denotes the maximum of $x$ and zero.
    \citet{BS_1973} and \citet{Merton} derive the replicating price $C(\cdot)$ of this call, assuming zero interest rate, as
    \begin{align*}
        C(t) &= \E[D|\CF(t)]\\
            &= S_1(t) \Phi\left(\frac{1}{\sqrt{1-t}}\left(\log(S_1(t)) + \frac{1-t}{2}\right)\right) -
            \Phi\left(\frac{1}{\sqrt{1-t}}\left(\log(S_1(t)) - \frac{1-t}{2}\right)\right)\\
            &\geq 0
    \end{align*}
    for all $t \in [0,1]$,
    where $\Phi$ denotes the standard normal cumulative distribution function.

    We now set
    \begin{align*}
        M(t) =  \int_0^t \1_{\{\varrho>s\}} \frac{1}{\sqrt{1-s}} \dd B(s)
    \end{align*}
    for all $t \in [0,1]$, where
    \begin{align}  \label{varrho}
        \varrho := \inf \left\{t \in [0,1]: \int_0^t \frac{1}{\sqrt{1-s}} \dd B(s) = C(0) + 1 \right\}.
    \end{align}
    Then, we have $\varrho < 1$, which yields $M(1) = C(0) + 1$.  This holds since the integral appearing in \eqref{varrho} is
    a continuous local martingale with quadratic variation $-\log(1-t)$, which tends to infinity as $t$ tends to one.
    Thus, it can be represented as time-changed Brownian motion, which almost surely hits $C(0) + 1$.

    We introduce a second asset with a price process $S_2(\cdot)$ specified as
    \begin{align}  \label{E S2}
        S_2(t) = C(t \wedge 1) + M(t \wedge 1) - C(0) - 1
    \end{align}
    for all $t \geq 0$, where $x \wedge y$ denotes the minimum of $x$ and $y$.
    We observe that $S_2(0) = -1$ and that $S_2(\cdot)$ is a local martingale that is
    neither bounded from above nor from below by a constant, but is a submartingale, as it is
    the sum of a martingale and a submartingale.
    Furthermore, and most importantly, $S_2(1) = C(1) = D$.  

    We now consider an economy consisting of a money market account paying zero interest rate and two assets with
    price processes given by  $S_1(\cdot)$ and $S_2(\cdot)$, as specified above.  We observe that this economy satisfies
    NFLVR, since both $S_1(\cdot)$ and $S_2(\cdot)$ are local martingales under the probability measure $\Prob$.
    Moreover, the second asset can be considered the price of a call written on the first asset with exercise price $1$,
    since its terminal payoff
    is exactly $D = (S_1(1) - 1)^+$.  In accordance to standard theory, we take exactly these trading strategies that lead to 		wealth processes bounded from below by a constant as    
     the class of admissible trading strategies.

    Any agent in this economy can replicate this call written on $S_1(1)$ for the price of $C(0) > 0>-1 = S_2(0)$.
    However, despite the existence of a market price for a call in the market, no arbitrage opportunity exists in this economy since the agent is not allowed to build a position that includes buying the call for a fixed time with price process $S_2(\cdot)$. To see this, consider the wealth process of a trading strategy that sells a portfolio that replicates the call with the dynamic Black-Scholes-Merton trading strategy, buys the second asset with price $S_2(0)$, and puts the profits of building this position in the money market account. The corresponding wealth process $W(\cdot)$ is thus the sum of three positions: a long position in $S_2(\cdot)$, a short position in the replicating portfolio,  and a holding in the money market. In other words, the wealth $W(t)$ at time $t \in [0,1]$ is exactly
    \begin{align*}
    	W(t) =  S_2(t) - C(t) + (C(0)-S_2(0)) = M(t).
	\end{align*}
	In particular, the initial wealth is zero, to wit, $W(0) = 0$, and the terminal wealth is strictly positive, to wit, $W(1) = C(0) + 1 >0$.  However, $W(\cdot)$ is not bounded from below as it is a time-changed (stopped) Brownian motion.  Thus, this trading strategy is not an arbitrage strategy since it is not admissible.
	Even if one extends the class of admissible trading strategies in the sense of \citet{Strasser2003}, this trading strategy is still not admissible as $W(\cdot)$ is not a supermartingale.

	Observe that we did not use any specific properties of call prices. Indeed, in \eqref{varrho} and \eqref{E S2}, we can replace $C(\cdot)$ by any nonnegative martingale $\widetilde{C}(\cdot)$ representing the minimal replicating cost of a contingent claim with payoff $\widetilde{C}(1)$ at time $1$. In this modified market, $S_2(\cdot)$ can then be interpreted as the price process of a contingent claim that pays $\widetilde{C}(1)$ at time $1$; observe that the initial price is again $S_2(0) = -1$.

    Admittedly, this example is quite pathological: It corresponds to an economy in which it is inadmissible to hold
    the second asset for a fixed deterministic time, although it is clearly admissible to hold the asset until a certain stopping time. However, this example also emphasizes that such pathological price processes as negative
    European call prices are not excluded by the NFLVR assumption.  Thus, any no-arbitrage argument based on
    constructing a trading strategy must ensure that this trading strategy is admissible. To illustrate, the
    standard argument that a European call price for a strike $K$ is bounded from below by $S_1(0) - K$ is often
    formulated as follows: Assume that the call price is smaller than $S_1(0) - K$. Then, consider the following
    trading strategy:
    Buy the call, sell the stock, borrow $K$ dollars and put the leftover money in the bank account. At maturity,
    this trading strategy has corresponding wealth of at least the positive amount of money in the bank account and thus
    seems to imply the existence of an arbitrage opportunity.
    However, in our example above, this trading strategy would already be inadmissible.

	\section{Concluding remarks}
    It is important to emphasize that the discussion so far only involved European-style contingent claims.  For example,
    American calls being in-the-money cannot have negative prices under the NFLVR
    assumption. An agent could buy such an American-style contingent claim and immediately exercise it, collecting at least the contingent claim's negative
    price.  \citet{BKX} observe that put-call parity holds, as long as the European
    call price is exchanged by the corresponding American call price; however, they also (explicitly)
    assume that both the European put price and the American call price are the corresponding replicating prices. In the same manner as above, it is easy to construct an arbitrage-free economy where put-call parity does not hold,
    even after replacing the European by the American call.

    The discussion in \citet{MadanYor_Ito} is related to different arbitrage arguments that can be made
    with respect to American and European-style contingent claims; they discuss, in the context of bubbles,
    robustness of trading strategies with respect to ``random early liquidations.''  For example, in the economy above, consider the two trading strategies
	of selling the call with corresponding  wealth process $W_1(t) = S_2(0) - S_2(t) = -1-S_2(t)$ and of selling the Black-Scholes-Merton replicating portfolio with corresponding wealth process $W_2(t) =C(0) - C(t)$ for all $t \in [0,1]$.
	Observe that $W_1(0) = 0 = W_2(0)$ and $W_1(T) < W_2(T)$.
	Both trading strategies are admissible under the weak admissibility constraints of  \citet{Strasser2003}. The second trading strategy seems better than the first one as it leads to a higher terminal wealth. 
    However, if an agent has to cover a short position in the call $S_2(\cdot)$ and bears the risk that her counterparty might liquidate this short position at some time $t \in (0,1)$, she cannot follow the trading strategy of replicating the call's terminal payoff since the wealth process $W_2(t)$
    can be (unboundedly far) below $W_1(t)$.
    
    Similarly, \citet{CH} discuss collateral requirements for European-style contingent claims. If an agent followed the
    Black-Scholes-Merton trading strategy to obtain a terminal wealth of $C(0)-C(1)$, her wealth process
    might not satisfy such a collateral requirement, which is basically an American-style
    feature and forces one's wealth process to stay above a certain barrier that, in this case, depends on the
    call price $S_2(\cdot)$.

    We have illustrated that simple no-arbitrage arguments rely on more assumptions than only on the existence of an
    ELMM.   Even if an agent observed negative European call prices quoted according to the example above, she could not achieve a nonnegative
    and with positive probability positive wealth at a later time by starting from zero initial wealth and following an admissible trading
    strategy.
From the purely economic perspective of equilibrium pricing, the above example is of little insight. 
No agent
    in that economy is allowed to hold the call for a fixed time. However, standard no-arbitrage proofs do not
    include this point in their argument.

    What assumptions do simple no-arbitrage arguments, relying on selling and buying certain assets, implicitly make?
    This question can be addressed in several ways. A technical assumption could be to consider only assets whose
    price processes are true martingales under a fixed ELMM.  Then, both buying and selling these assets
    (and a combination of buying and selling) yield
    admissible trading strategies. 

An assumption in more economic terms is the \emph{no-dominance} principle,
    as suggested by \citet{Merton}, which is a slightly stronger assumption than NFLVR.
    The no-dominance principle basically states that if trading strategy A leads to a wealth greater than or equal to
    the wealth of trading strategy B, then the initial cost of trading according to A should be greater than or equal to
    the initial cost of trading according to B.
    For instance, if no dominance holds, then European call prices on a nonnegative stock price have to be nonnegative.
    To see this, compare the trading strategy of holding the call to the trading strategy of doing nothing at all, costing
    zero and leading to a terminal wealth of zero, which is less than or equal to the terminal wealth corresponding to
    holding the call. Thus, the no-dominance principle yields that any call price has to be nonnegative. For details on the no-dominance principle and for a study how far this additional assumption can take us, see \citet{JP_incomplete}.
\bibliography{aa_bib}{}
\bibliographystyle{apalike}

\end{document}